# Two-Fold-Symmetric Magnetoresistance in Single Crystals of Tetragonal BiCh$_2$-Based Superconductor LaO$_{0.5}$F$_{0.5}$BiSSe

Kazuhisa Hoshi[1], Motoi Kimata[2], Yosuke Goto[1], Tatsuma D Matsuda[1], Yoshikazu Mizuguchi[1]

1 *Department of Physics, Tokyo Metropolitan University, 1-1, Minami-osawa, Hachioji 192-0397, Japan*
2 *High Field Laboratory for Superconducting Materials, Institute for Materials Research, Tohoku University, Sendai 980-8577, Japan*

Abstract

We have investigated the in-plane anisotropy of the *c*-axis magnetoresistance for single crystals of a BiCh$_2$-based superconductor LaO$_{0.5}$F$_{0.5}$BiSSe under in-plane magnetic fields. We observed two-fold symmetry in the *c*-axis magnetoresistance in the *ab*-plane of LaO$_{0.5}$F$_{0.5}$BiSSe while the crystal possessed a tetragonal square plane with four-fold symmetry. The observed symmetry lowering in magnetoresistance from the structural symmetry may be related to the nematic states, which have been observed in the superconducting states of several unconventional superconductors.

The BiCh$_2$-based (Ch: S, Se) superconductor family was discovered in 2012 [1–3], and it has been extensively studied because of the layered crystal structure similar to those of the cuprate and iron-based high-transition-temperature (high-$T_c$) superconductors: all those layered superconductors have a square network of conduction planes [4,5]. The typical REOBiCh$_2$ (RE: La, Ce, Pr, Nd) system has a layered structure composed of alternate stacks of REO insulating layers and BiCh$_2$ conducting layers [2,3]. F substitutions at the O site generate electron carriers in the BiCh$_2$ layers, which results in the emergence of metallicity and superconductivity. The highest $T_c$ among the BiCh$_2$-based superconductors is 11 K for LaO$_{0.5}$F$_{0.5}$BiS$_2$ prepared under high pressure [6,7]. Although the material variation has been widely developed, the pairing mechanisms of the superconductivity of BiCh$_2$-based superconductors have not been completely clarified. Recent theoretical calculations and angle-resolved photoemission spectroscopy (ARPES), however, have suggested that unconventional mechanisms are essential for the superconductivity of BiCh$_2$-based superconductors [8,9]. In the ARPES study, an anisotropic superconducting gap with nodes was observed in Nd(O,F)BiS$_2$ [9]. In addition, from a study of the selenium isotope effect on a BiCh$_2$-based superconductor LaO$_{0.6}$F$_{0.4}$BiSSe, the isotope effect exponent $\alpha$ was found to be close to zero (-0.04 < $\alpha$ < +0.04) [10]. This is inconsistent with the conventional electron-phonon mechanism, which generally gives $\alpha \sim 0.5$.

Recently, for layered superconductors, *nematicity* (or *nematic* state) has been a hot topic owing to the its possible relation to the unconventional superconductivity states. Nematicity has been originally used for liquid crystals, and nematicity in a layered compound can be described as the spontaneous unidirectional state in which rotational symmetry breaking (lowering) is observed. Electronical nematicity has been observed in various unconventional superconductors such as cuprate and iron-based superconductors. Those superconductors show a rotational symmetry breaking of the electronic structure while the original (four-fold) lattice symmetry is maintained [11,12]. Furthermore, nematic superconductivity has been observed in Cu$_x$Bi$_2$Se$_3$ [13–15]. The Cu$_x$Bi$_2$Se$_3$ superconductor is known as an odd-parity superconductor, and the odd-parity states can be regarded as topological superconductivity states [14]. In the superconducting states of Cu$_x$Bi$_2$Se$_3$, two-fold symmetry of a superconducting gap was observed while the conducting layer possessed a hexagonal network. In addition, similar symmetry breaking in the superconducting state was observed in Sr$_x$Bi$_2$Se$_3$ as well [16]. Moreover, the superconductivity in Sr$_2$RuO$_4$, which is widely considered to be a chiral $p$ wave, has been proposed as the nematic superconductivity state [17].

Since recent studies on BiCh$_2$-based compounds have proposed unconventional pairing mechanisms, we expected the observation of nematic or related exotic phenomena in the superconducting states of BiCh$_2$-based compounds. In this study, we investigated the in-plane anisotropy of the $c$-axis magnetoresistance of LaO$_{0.5}$F$_{0.5}$BiSSe single crystals under in-plane magnetic fields. From in-plane anisotropy measurements on the $c$-axis magnetoresistance, we observed the appearance of two-fold symmetry with in-plane azimuth field angles while the crystal structure (the conducting plane) possessed a tetragonal square plane with four-fold symmetry.

$LaO_{0.5}F_{0.5}BiSSe$ single crystals were grown by using a high-temperature flux method in an evacuated quartz tube. First, polycrystalline samples of $LaO_{0.5}F_{0.5}BiSSe$ were prepared by the solid-state-reaction method using powders of $La_2O_3$ (99.9%), $La_2S_3$ (99.9%), $Bi_2O_3$ (99.999%), and $BiF_3$ (99.9%) and grains of Bi (99.999%) and Se (99.999%), as described in Ref. 18. A mixture of the starting materials with a nominal ratio of $LaO_{0.5}F_{0.5}BiSSe$ was mixed, pressed into a pellet, and annealed at 700 ºC for 20 h in an evacuated quartz tube. The polycrystalline powder of $LaO_{0.5}F_{0.5}BiSSe$ (0.62 g) was mixed with CsCl flux (2.2 g), and the mixture was sealed into an evacuated quartz tube. The tube was heated at 900 ºC for 12 h and slowly cooled to 645 ºC at a rate of -1.0 ºC /h, followed by furnace cooling to room temperature. At room temperature, the quartz tube was opened under air atmosphere, and the product was filtered and washed with pure water. The single crystals were analyzed by scanning electron microscopy (SEM). As shown in the inset of Fig. 1(a), plate-like crystals were obtained. The plate surface with a square shape corresponds to the $ab$-plane of tetragonal $LaO_{0.5}F_{0.5}BiSSe$. The chemical composition of the obtained crystal was investigated by energy-dispersive X-ray spectroscopy (EDX). The average ratio of the constituent elements (except for O and F) was estimated to be La : Bi : S : Se = 1 : 0.98 : 1 : 0.98, which was normalized by the La value. The analyzed atomic ratio is almost consistent with the nominal composition of $LaO_{0.5}F_{0.5}BiSSe$. Considering the error in the EDX analysis, we regard the S : Se composition of the examined crystal as 1 : 1.

Powder X-ray diffraction (XRD) experiment was performed at room temperature by using RIGAKU Miniflex600 equipped with a D/teX-Ultra detector with a Cu-K$\alpha$ radiation. Figure 1(a) shows the powder XRD pattern of the $LaO_{0.5}F_{0.5}BiSSe$ single crystals. Only 00$l$ peaks were observed, which confirms that the $ab$-plane is well developed. The $c$-axis lattice constant is roughly estimated as 13.565(7) Å. Figure 1(b) shows the F concentration dependences of the $c$-axis lattice constant for the polycrystalline and single-crystal samples of $LaO_{1-x}F_xBiSSe$. Figure 1 (b) shows the F concentration dependences of the $c$-axis lattice constant for the polycrystalline (open rectangles [18]) and single-crystal (black diamond [23], blue triangle[24], and red circles (this work)) samples of $LaO_{1-x}F_xBiSSe$. The estimated $c$ for the present crystal is close to that for the polycrystalline sample with $x = 0.5$ [18]. Therefore, in this study, we call the examined crystal $LaO_{0.5}F_{0.5}BiSSe$ using the nominal composition. In addition, we performed single-crystal X-ray structure analysis at room temperature on a $LaO_{0.5}F_{0.5}BiSSe$ single crystal taken from the same batch of crystals used in this study. The structural parameters were refined with the tetragonal ($P4/nmm$) structural model using the refinement program of SHELXL [19]. The analysis results are summarized in Tables S1 and S2 (Supplemental Materials [20]).

The magnetoresistance measurement was performed using a superconducting magnet and the 25 T cryogen-free superconducting magnet [21] at the high field laboratory of the Institute for Materials Research (IMR), Tohoku University. Figure S1 (Supplemental Materials [20]) shows the field dependence of the electrical resistance and the temperature dependence of the $H_{c2}$ for our single crystal of LaO$_{0.5}$F$_{0.5}$BiSSe. The $H_{c2}$ ($H//ab$) obtained from the Werthamer-Helfand-Hohenberg (WHH) model fitting [22] was 29.5 T for our single crystal. The large in-plane $H_{c2}$ is consistent with the results of previous reports [23,24]. The extremely high $H_{c2}$ could be originating from the breaking of local inversion symmetry at the conducting layers in the BiCh$_2$-based superconductors [25]. Hence, we used a high field for investigating the in-plane anisotropy of $H_{c2}$. To control the magnetic field direction precisely, a two-axis rotational probe was used. Figure 2(a) shows a schematic image of the terminal configuration (made using Au wires and Ag pastes) for the $c$-axis resistivity measurements. The angles $\theta$ and $\phi$ are defined as show in Fig. 3 (a). The angle $\theta$ is measured from the $c$-axis, and $\phi$ is the azimuth angle measured from the $a$-axis to the $b$-axis, although the $a$-axis and the $b$-axis are equivalent in a tetragonal crystal. Note that the charge current is always perpendicular to the magnetic field when the magnetic field is in the conducting plane, since the $c$-axis resistivity was measured. Figure 2(b) shows the temperature dependence of the $c$-axis electrical resistivity for LaO$_{0.5}$F$_{0.5}$BiSSe single crystal. The $T_c^{\text{onset}}$ is 4.3 K, and $T_c^{\text{zero}}$ is 3.7 K.

Figure 3(c) shows the $\theta$ angle dependence of the $c$-axis resistivity at the conditions of $\phi = 90°$, $\mu_0 H = 15$ T, and $T = 2.5$ K. $\rho_{\min}$ is defined as the minimum $c$-axis resistivity at which the magnetic field is exactly parallel to the conducting plane. Figure 3(d) shows the $\phi$ dependences of $\rho_{\min}$ at several temperatures in the range 2.0–5.0 K for Sample A shown in Fig. 3 (b). Below 3.0 K, the $\phi$ dependence of $\rho_{\min}$ shows dips at $\phi = \pm 90°$, indicating two-fold symmetry of $\rho_{\min}$. With decreasing temperature, the amplitude of the two-fold-symmetric $\rho_{\min}(\phi)$ becomes apparent. At 2.0 K, the two-fold symmetry of $\rho_{\min}(\phi)$ is still observed while $\rho_{\min}$ approaches zero resistivity; the noisy data is due to the low resistivity close to our measurement limit. The observation of the two-fold symmetry of $\rho_{\min}(\phi)$ in the superconducting states was unexpected because the crystal structure (the conducting plane) possesses a tetragonal (four-fold) symmetry. Furthermore, the two-fold symmetry was observed in the $\phi$ dependence of $\rho_{\min}$ for another sample (Sample B in Fig. 3 (b)) obtained from a different batch (See Fig. S2 of Supplemental Materials [20]). We confirmed that the two-fold symmetric behavior in $\rho_{\min}$ is similarly observed when the electrode configuration is changed (rotated by 90°) in the distinct sample. Therefore, the two-fold symmetry in $\rho_{\min}(\phi)$ is not originating from the possible inhomogeneous current flow (See Fig. 3(b)). At 5.0 K, $\rho_{\min}$ is almost independent of $\phi$, which suggests that the observed two-fold symmetry appears in superconducting states but not in normal states.

Now, we discuss the possible origins of the observed two-fold symmetry of $\rho_{\min}(\phi)$ in the superconducting states of LaO$_{0.5}$F$_{0.5}$BiSSe. First, it should be noted that there is a possibility of symmetry lowering of the conducting plane structure because some of the parent compounds of BiCh$_2$-based superconductors show structural instability and a structural transition from tetragonal (*P*4/*nmm*) to monoclinic (*P*2$_1$/*m*). However, the structural transition is typical for parent (non-doped) phases and is suppressed by carrier doping. In fact, our single-crystal structural analysis showed that the space group of our LaO$_{0.5}$F$_{0.5}$BiSSe crystal at room temperature was tetragonal, which is consistent with the results of previous studies [18,23]. In addition, as shown in Fig. S4 of Supplemental Materials [20], the temperature scan of the synchrotron powder X-ray diffraction pattern for the polycrystalline sample of LaO$_{0.5}$F$_{0.5}$BiSSe (not single crystal) confirmed no structural transition down to 100 K. The split of the 200 peak was observed when the transition from tetragonal to monoclinic (or orthorhombic) occurred. Furthermore, no anomaly was observed in the temperature dependence of resistivity from 300 to 5 K (See Fig. S3 of Supplemental Materials [20]). Therefore, we consider that the observed two-fold symmetry is due to electronic (or orbital) origins and not due to structural origins. Then, the possible scenario is the appearance of nematic states as observed in other unconventional layered superconductors such as the cuprate, Fe-based, and doped Bi$_2$Se$_3$ superconductors [13–16]. Although there has been no theoretical prediction about nematicity in the superconducting states of BiCh$_2$-based compounds, we can expect that unconventional mechanisms of superconductivity are emerging in the examined crystal, on the basis of recent theoretical and experimental studies on the BiCh$_2$-based system [8–10]. Hence, the observed two-fold symmetry of the *c*-axis magnetoresistance may be related to the unconventional superconductivity linked with fluctuations of electronic states. To obtain further knowledge on the superconductivity mechanisms and the anisotropic (possibly nematic) states, studies of anisotropy on thermodynamic parameters for BiCh$_2$-based superconductors are desired.

In conclusion, we have investigated the in-plane anisotropy of the *c*-axis magnetoresistance of single crystals of a BiCh$_2$-based superconductor LaO$_{0.5}$F$_{0.5}$BiSSe under in-plane field configuration. We observed two-fold symmetry of the *c*-axis magnetoresistance in the *ab* plane of LaO$_{0.5}$F$_{0.5}$BiSSe while the conducting plane possessed a tetragonal square plane with four-fold symmetry. We consider that the observed in-plane anisotropy of the *c*-axis magnetoresistance would be due to electronic (or orbital) origins and not due to structural origins. The observed two-fold symmetry is possibly related to the nematic states, which have been observed in the superconducting states of several unconventional superconductors with a layered structure.

Acknowledgement

We would like to thank C. Moriyoshi, Y. Kuroiwa, A. Miura, and O. Miura for their experimental support and fruitful discussions. This work was partly supported by Collaborative Research with IMR, Tohoku Univ. (proposal number: 17H0074) and Grants-in-Aid for Scientific Research (Nos. 15H05886, 16H04493, 17K19058).

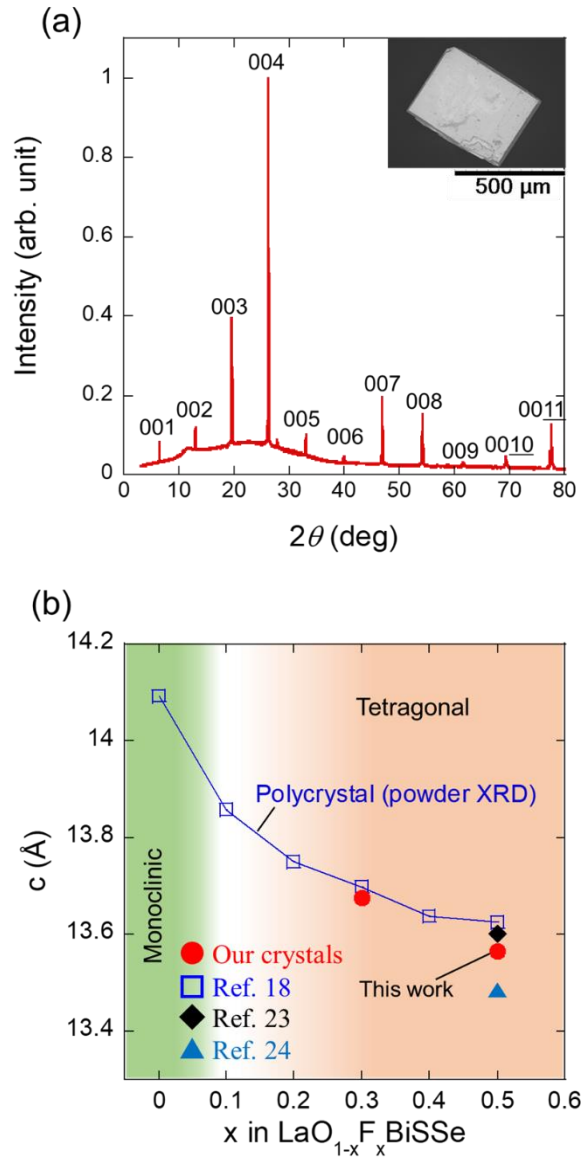

Fig. 1. (color online) (a) XRD pattern of $LaO_{0.5}F_{0.5}BiSSe$ single crystals. The inset shows an SEM image of a $LaO_{0.5}F_{0.5}BiSSe$ single crystal. (b) F concentration ($x$) dependences of the $c$-axis lattice constant for single crystals and polycrystalline samples of $LaO_{0.5}F_{0.5}BiSSe$ [18, 23, 24].

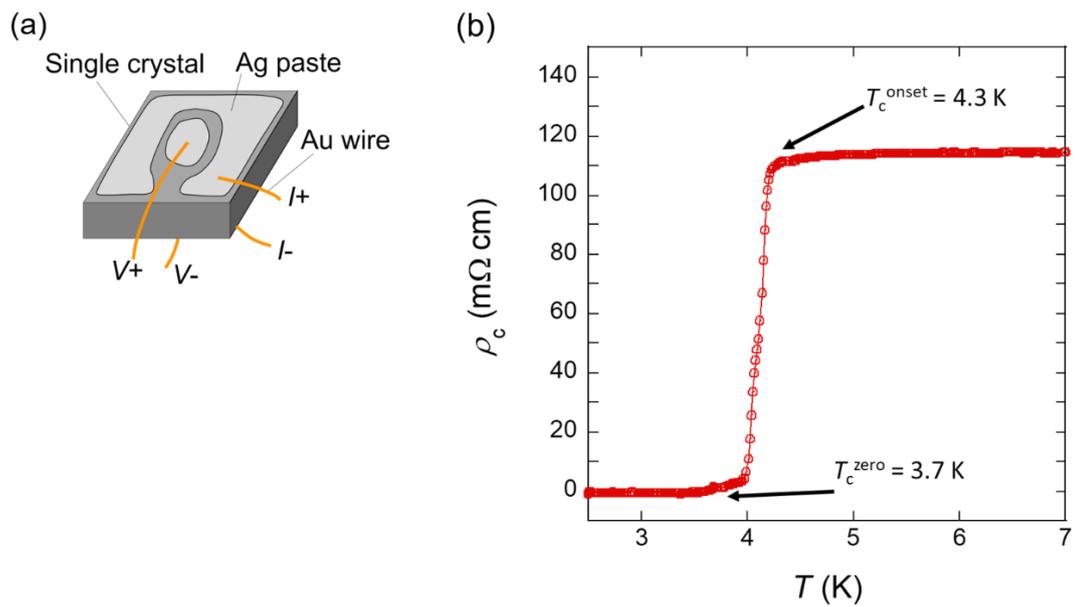

Fig. 2. (color online) (a) Schematic image of the terminal configuration (Au wires and Ag pastes) for the *c*-axis electrical resistivity measurements. (b) Temperature dependence of the *c*-axis electrical resistivity for a LaO$_{0.5}$F$_{0.5}$BiSSe single crystal from 2.5 to 7.0 K.

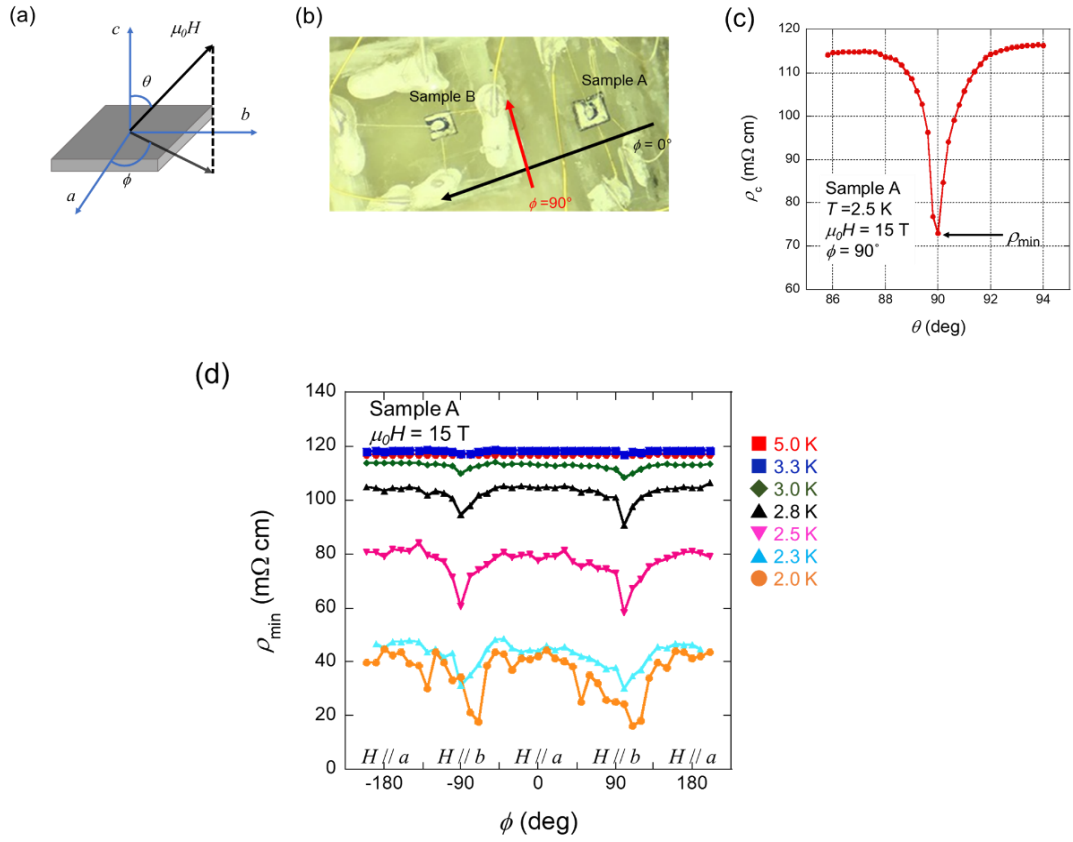

Fig. 3. (color online) (a) Schematic image of the rotation angles for the anisotropy measurement. (b) Photograph of samples and terminals used in the measurements. The in-plane anisotropy measurements of sample A and B are shown in the Fig. 3 (d) and Fig. S2 (Supplemental Materials [20]), respectively. (c) $\theta$ angle dependence of the $c$-axis electrical resistivity at $\phi = 90º$, $\mu_0 H = 15$ T, and $T = 2.5$ K. (d) $\phi$ dependences of $\rho_{min}$ at several temperatures in the range 2.0-5.0 K.

[Supplemental Materials]

**Two-Fold-Symmetric Magnetoresistance in Single Crystals of Tetragonal BiCh$_2$-Based Superconductor LaO$_{0.5}$F$_{0.5}$BiSSe**

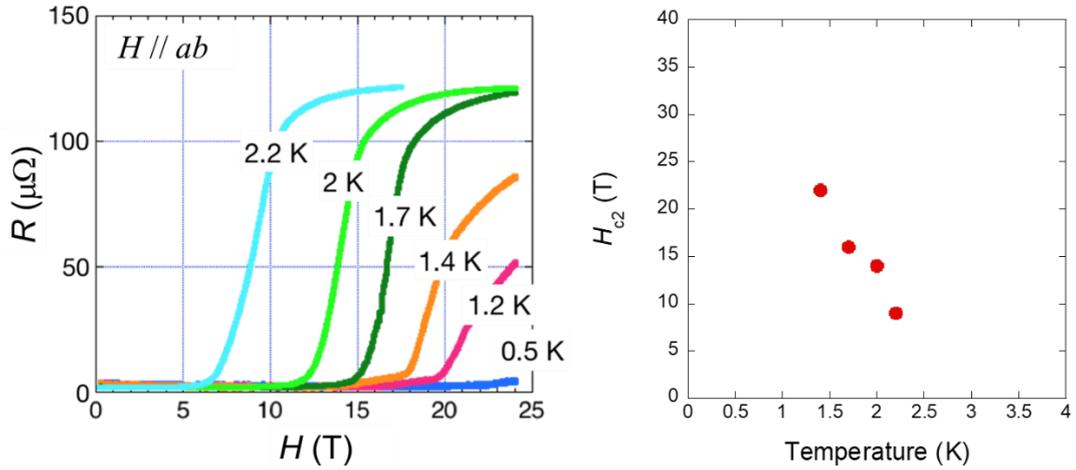

Fig. S1(a). The field dependence of the electrical resistance with various temperature for our single crystal of LaO$_{0.5}$F$_{0.5}$BiSSe obtained from the same batch of the investigated crystal. Fig. S1(b). The temperature dependence of the $H_{c2}$. The $H_{c2}$ is defined as the mid-points at which the resistance drop is half of normal state.

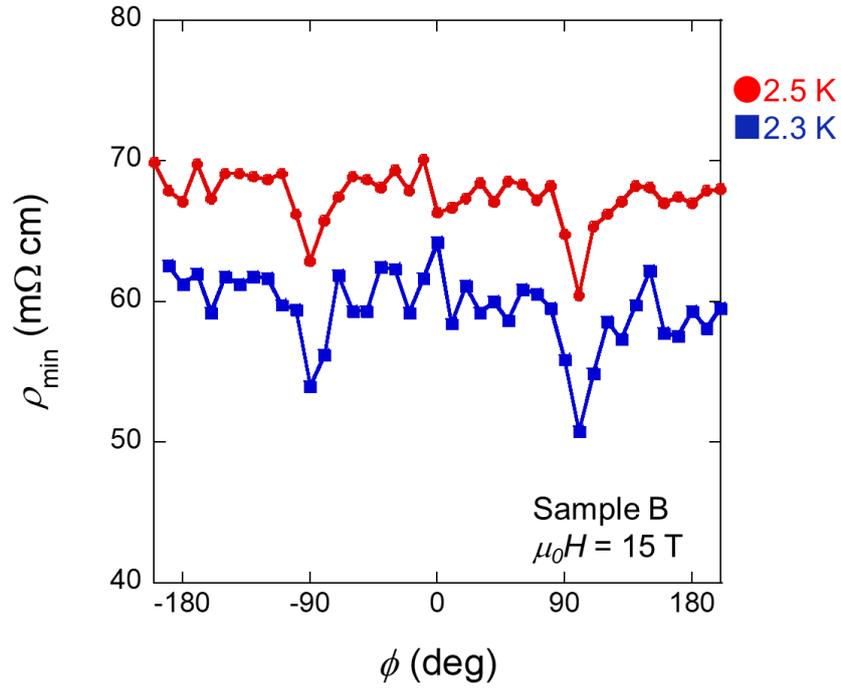

Fig. S2. $\phi$ angle dependences of the $\rho_{min}$ at $T$ = 2.3 K and 2.5 K for a crystal of LaO$_{0.5}$F$_{0.5}$BiSSe obtained from a different batch (Sample B in Fig. 3 (b)).

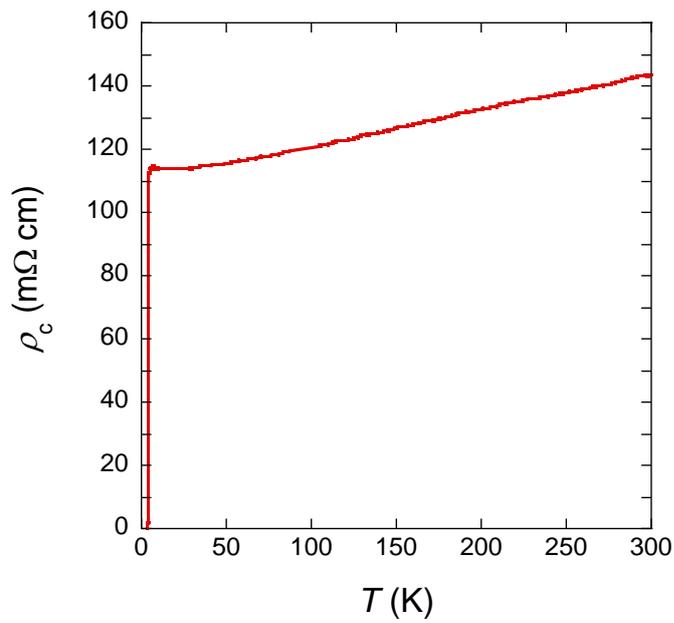

Fig. S3. Temperature dependence of electrical resistivity for the LaO$_{0.5}$F$_{0.5}$BiSSe crystal.

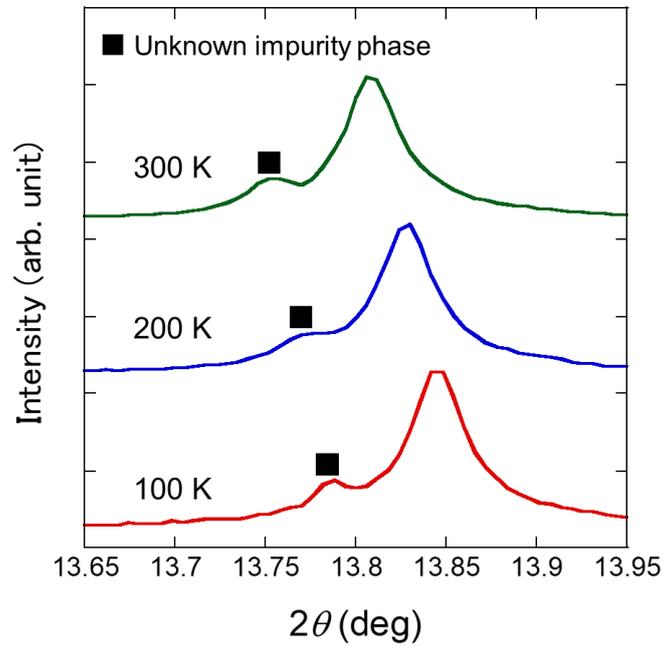

Fig. S4. The temperature dependence of the 200 peak of powder synchrotron X-ray diffraction for the polycrystalline sample of LaO$_{0.5}$F$_{0.5}$BiSSe. A structural transition from tetragonal to lower-symmetry (monoclinic or orthorhombic) was not observed at $T > 100$ K. The data was corrected with synchrotron X-ray [$\lambda = 0.495586(1)$ Å] at BL02B2, SPring-8 under a proposal of No. 2017B1211.

Table S1. Refined atomic coordinates and $B_{eq}$.

| site | x | y | z | $B_{eq}$ (Å²) |
|---|---|---|---|---|
| Bi | 0.25000 | 0.25000 | 0.87293(6) | 1.22(3) |
| La | - 0.25000 | - 0.25000 | 0.59611(9) | 0.77(3) |
| Ch1 (Se) | - 0.25000 | - 0.25000 | 0.87578(19) | 1.50(5) |
| Ch2 (S) | 0.25000 | 0.25000 | 0.6861(3) | 0.36(7) |
| O/F | - 0.25000 | 0.25000 | 0.50000 | 0.35(19) |

Table S2. Refined structural parameters and analysis condition.

| Formula | LaOBiSSe |
|---|---|
| Formula weight | 474.91 |
| Space group | Tetragonal $P4/nmm$ (#129) |
| Lattice type | Primitive |
| Z value | 2 |
| a (Å) | 4.1348(6) |
| c (Å) | 13.569(3) |
| V (Å³) | 231.98(7) |
| R | 0.0280 |
| $wR_2$ | 0.0808 |
| Crystal dimensions | 0.100 × 0.080 × 0.040 mm |
| Diffractometer | XtaLAB mini (RIGAKU) |
| Radiation | MoK$_\alpha$ ($\lambda$ = 0.71075 Å) (graphite monochromated) |
| Temperature (K) | 293 |